\documentclass[conference]{IEEEtran}

\usepackage{amsmath,amssymb,graphicx,xcolor,hyperref,fancyhdr}
\usepackage{booktabs}
\usepackage{hyperref}
\usepackage{array}
\definecolor{metablue}{RGB}{0,141,169}

\hypersetup{
    colorlinks=true,
    linkcolor=metablue,
    citecolor=metablue,
    urlcolor=metablue
}

\makeatletter
\fancypagestyle{plain}{
    \fancyhf{}
    \fancyhead[C]{META 2026, DUBLIN -- IRELAND, JULY 14 -- 17, 2026}
    
}
\fancypagestyle{IEEEtitlepagestyle}{
    \fancyhf{}
    \fancyhead[C]{META 2026, DUBLIN -- IRELAND, JULY 14 -- 17, 2026}
    
}
\makeatother

\title{Symmetry-Protected Quantum Computing\\ using Metamaterials}

\author{
N. F. Johnson\textsuperscript{1,*}, F. J. Rodr\'{i}guez\textsuperscript{2}, L. Quiroga\textsuperscript{2} \\[1ex]
\textsuperscript{1}Physics Department, George Washington University, Washington, DC, USA \\
\textsuperscript{2}Departamento de F\'{i}sica, Universidad de Los Andes, Bogot\'a D.C., Colombia \\
*Corresponding author: \href{mailto:neiljohnson@gwu.edu}{neiljohnson@gwu.edu}
}

\begin{document}
\maketitle

\begin{abstract}
We propose a new architecture for practical quantum computing that combines three established principles: symmetry protection of relative-motion qubits via the generalized Kohn theorem, control via twisted-light orbital angular momentum, and metamaterial nanofocusing (e.g. using Weyl-semimetal plasmonics). Crucially, the core mechanism is generic: it applies to any current or future quantum computing system involving parabolic confinement, including cold atoms, ions, and semiconductor dots.

\end{abstract}

\section{Introduction}

The generalized Kohn theorem~\cite{BJH1989,BJ1990,QJ1,QJ2}  is a remarkably general symmetry result whose implications are currently under-exploited for quantum information processing, including quantum computing. For any quantum system confined by a parabolic potential, the theorem~\cite{BJH1989,BJ1990,QJ1,QJ2} guarantees that the relative-motion degrees of freedom are exactly decoupled from any spatially uniform perturbation, regardless of its temporal character. Whether the perturbation is static drift, slow 1/$f$ fluctuation, GHz electromagnetic noise, or fast stochastic shot noise --- if the electric field is approximately uniform on the trap scale, it couples only to the center-of-mass coordinate and leaves the internal spectrum invariant. \emph{This protection is exact} rather than perturbative, holds for arbitrary interaction strength among the confined particles, and applies in the same form to semiconductor quantum dots, cold atoms in harmonic traps, ions in Paul traps, electrons on liquid helium, and any other quantum architecture built on harmonic confinement (see Fig. 1 schematic).

The relevance of this generality to practical quantum computing is greater than might first appear, because approximately parabolic confinement is not an unusual special case but a generic feature of physical quantum systems. By Gauss's law, a uniform charge distribution produces a linearly varying electric field whose integral is a parabolic potential; this is the foundation of Penning traps, gate-defined quantum dots, and the Fock--Darwin model~\cite{Reimann2002} that underlies most theoretical work on confined-electron systems. Depletion regions at junction edges in semiconductor heterostructures, lateral confinement set by surface gates above a two-dimensional electron gas~\cite{Hayashi2003,Petersson2010}, and radial confinement provided by RF quadrupole electrodes in ion traps~\cite{IonOAM2016} all produce approximately quadratic-in-displacement potentials over the relevant operating volumes. The Kohn theorem therefore applies, by construction, to a broad class of present-day and proposed quantum computing components, and the architectural principle developed here is significantly more general than its semiconductor instantiation.

We nonetheless focus on semiconductor implementations in this paper for two reasons. First, the long-term integration of quantum computing with artificial-intelligence (AI) workloads will require dense interfacing with classical accelerator hardware: semiconductor quantum-dot platforms~\cite{Friesen2017,Koski2020} have a clearer path to that integration than alternative platforms (cold atoms, ions), which currently face significant engineering obstacles. Second, the metamaterial-photonics research base --- including silicon-photonics infrastructure, GaAs heterostructure technology, and the emerging field of Weyl-semimetal device engineering~\cite{Guo2023} --- is concentrated in semiconductor materials. The architecture proposed here therefore sits at the intersection of two semiconductor-rooted research programs (quantum information processing and metamaterial photonics), while retaining the option to transfer to other parabolic-confinement platforms as needed.

\begin{figure*}[!t]
\centering
\includegraphics[width=0.60\textwidth]{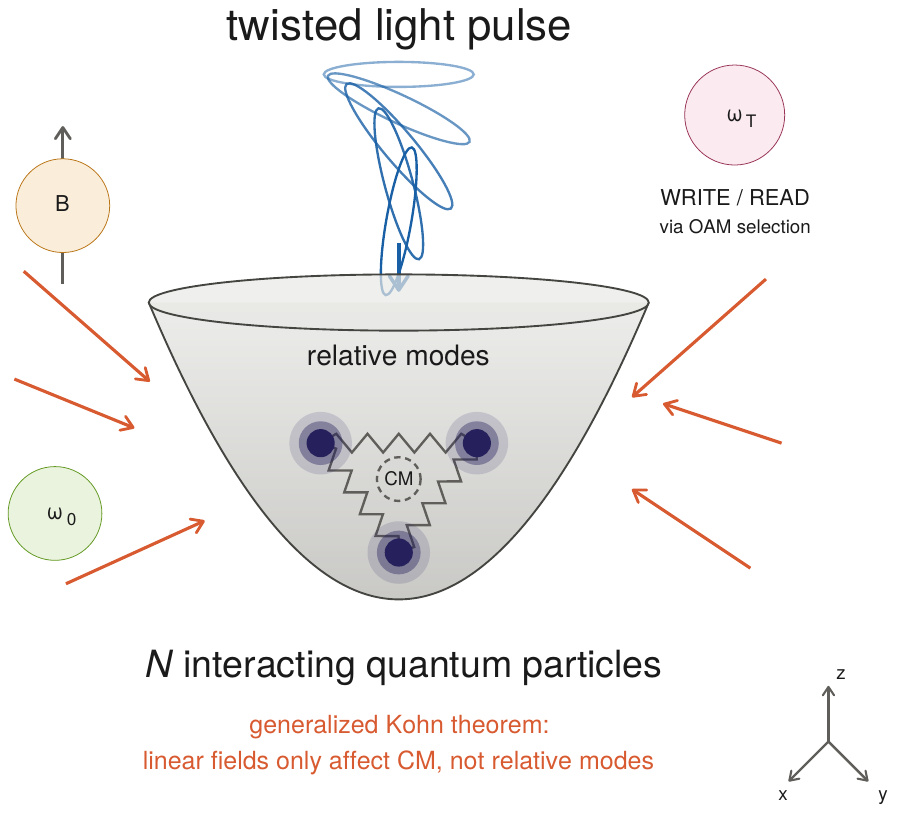}
\caption{Schematic of the symmetry-protected quantum-computing architecture. $N$ interacting quantum particles in a parabolic confinement (frequency $\omega_0$). A twisted-light pulse with orbital angular momentum at central frequency $\omega_T$ performs \textsc{Write} and \textsc{Read} operations through the OAM selection rule $\Delta|m|=\pm(l+\sigma)$. Examples of the quantum particles include, but are not limited to, electrons or holes in a semiconductor system. A perpendicular magnetic field $B$ can then be used to manipulate the relative-motion modes. Stray linear electric fields entering from any direction and at any temporal frequency couple only to the center-of-mass coordinate (generalized Kohn theorem) and therefore cannot dephase the relative-motion qubit. 
Recent work of Peluso et al.~\cite{Peluso2025} shows that surface plasmon
polaritons on a magnetic Weyl-semimetal conical tip can deliver
OAM-selective nanofocusing: all modes with a given sign of orbital angular
momentum can be focused at the end of the tip. This is in stark contrast with normal
metals, for which only one mode can reach the end. Combined with our proposal's
twisted-light selection rules, this provides a
near-field route for writing and reading correlation-sector qubits with
structured optical angular momentum.}
\label{fig:schematic}
\end{figure*}

A deployable quantum computing platform requires three capabilities in combination: a qubit encoding that survives environmental noise, a control channel that addresses the encoding with high fidelity, and a way to focus that control channel onto the qubit at the relevant length scales. Current platforms achieve each of these in isolation but rarely all three through a unified physical principle. This work proposes such a principle, drawing together three previously distinct bodies of work --- the generalized Kohn theorem for parabolic confinement, twisted-light orbital-angular-momentum (OAM) control of correlated electrons~\cite{Rodriguez2025,APSOS2026}, and OAM-selective metamaterial nanofocusing~\cite{Peluso2025} --- into a coherent architecture for practical quantum information processing. Figure~\ref{fig:schematic} summarizes the proposal.

\subsection{Charge-noise decoherence and the limits of existing protection}

The dominant decoherence channel for semiconductor-based qubits is fluctuating electric fields, encompassing 1/$f$ charge noise from two-level fluctuators in oxide layers and interfaces, gate-voltage drift through control wiring, and stray impurity charges polarizing on slow timescales~\cite{Gamble2012}. Because the typical distance from the dot to such sources is significantly larger than the dot itself ($\sim 10$--$100$~nm), the resulting electric field is well approximated as spatially uniform and dipolar across the qubit volume. Semiconductor charge qubits in GaAs exhibit $T_2^\ast \sim 1$~ns at generic operating points~\cite{Hayashi2003}, reaching $\sim 7$~ns only at carefully engineered ``sweet spots'' where the leading sensitivity vanishes by construction~\cite{Petersson2010}.

Existing protection strategies fall into three classes. {\em Sweet-spot operation} achieves first-order insensitivity at isolated parameter-space points: singlet--triplet qubits exhibit a sweet spot at the symmetric operating point with $T_2^\ast$ extended by two orders of magnitude~\cite{Petta2005}, exploited via symmetric exchange operation~\cite{Reed2016,Martins2016}. The limitation is that protection holds at the sweet spot but not along full gate trajectories. {\em Decoherence-free subspaces} (DFS), originally developed for spin qubits~\cite{ZanardiRasetti1997,LidarWhaley2003,BBW2000,KempeDFS2001}, encode logical states such that environmental noise acts identically on both, producing a global phase rather than dephasing. The most direct application to charge qubits is the {\em charge quadrupole qubit} of Friesen et al.~\cite{Friesen2017}: a single electron in a linear triple quantum dot is encoded such that uniform electric field fluctuations couple equally to both logical states. Predicted gate fidelities are 10--1000 times better than conventional charge qubits. This scheme has been experimentally realized by Koski et al.~\cite{Koski2020}, who demonstrated strong coupling between a superconducting microwave resonator and the electron quadrupole moment in GaAs/AlGaAs at an operating point where dipole coupling vanishes by geometric symmetry, with residual coherence limited by short-range (quadrupolar) charge noise. Subsequent work has continued the development~\cite{Kornich2018,Kratochwil2021,Russ2018}. {\em Parameter-regime protection} as exemplified by superconducting transmons (charge dispersion exponentially small at large $E_J/E_C$) and {\em matrix-element protection} as in spin qubits (no first-order spin--electric coupling) round out the established taxonomy of approaches.

A common feature of these approaches is that protection emerges from a specific implementation: a tuned operating point, a geometric arrangement of dots, a parameter regime, or an atomic-structure choice. None offers a {\em symmetry-based} protection of a many-body encoding that applies generically across implementation platforms.

\subsection{Twisted light as a quantum control channel}

Light carrying orbital angular momentum~\cite{Allen1992} has been studied as a probe of confined electronic states for over a decade. Quinteiro and Berakdar~\cite{Quinteiro2009Rings} analyzed OAM-induced electric currents in quantum rings; Quinteiro and Tamborenea~\cite{QuinteiroTamb2009Dots,QuinteiroTamb2009EPL} derived light--matter matrix elements for disc-shaped quantum dots and identified the OAM-conservation selection rule; subsequent work extended the analysis to bulk semiconductors~\cite{Sbierski2015}, off-centered beams~\cite{QuinteiroLuceroTamb2010}, strong-magnetic-component twisted-light regimes~\cite{QuinteiroReiterKuhn2017}, and magnetic-optical transitions~\cite{QuinteiroReiterKuhn2017Magnetic}. W\"atzel and Berakdar~\cite{Watzel2017} demonstrated structured-light photovoltaic effects. Together, such advances help explain why twisted light has evolved as a central resource in high-dimensional optical information processing \cite{Andrei,YaoPadgett2011,Wang2012,Krenn2016,Willner2021,Erhard2018,Norden2025} -- beyond its role as a purely spectroscopic and control probe.

The conceptual role of twisted light in quantum information has emerged more recently. Grass et al.~\cite{Grass2018} showed that nonzero-OAM light can pump angular momentum into a fractional quantum Hall droplet to create quasiholes and impose Berry phases on their motion. 
Macaluso et al.~\cite{Macaluso2020} numerically characterized the fractional charge and anyonic statistics of lattice quasiholes from density profiles in a Harper--Hofstadter fractional Chern-insulator setting.
Beterov et al.~\cite{Beterov2023} proposed driving forbidden atomic clock transitions with twisted light. Lee, Kim, and Lim~\cite{LeeKim2015} considered OAM control of superconducting qubits without Josephson junctions. The cold-atom community has parallel infrastructure: Léonard et al.~\cite{Leonard2023} have realized a two-particle Laughlin-type FQH state in a $4\times 4$ optical lattice; Lunt et al.  ~\cite{36point5} produced a Laughlin state of two rapidly rotating fermions; and Umucalilar~\cite{Umucalilar2023} has proposed a minimal cold-atom quasihole protocol. Earlier theoretical work~\cite{Popp2004,Grusdt2014,Barkeshli2015,Repellin2017,Lacki2025} mapped out FQH preparation paths in cold atoms.

These efforts establish that twisted light couples to internal states of confined electronic systems and obeys angular-momentum selection rules. What has been missing is the explicit recasting of these results in the language of qubit encoding, gate design, and noise immunity. Our recent work helps close this gap: Rodr{\'i}guez, Quiroga, and Johnson~\cite{Rodriguez2025} derived exact twisted-light matrix elements and chiral selection rules for interacting electrons in magnetized parabolic quantum dots in the analytically solvable Calogero limit; the subsequent paper~\cite{APSOS2026} recast these results as a concrete control primitive for quantum computing, defining a correlation-sector qubit encoding, mapping to a universal single-qubit gate set, and proposing a two-qubit entangling mechanism via state-dependent Coulomb coupling. The present paper extends that program by identifying the metamaterial-photonics layer required for a practical implementation.

\subsection{Metamaterial nanofocusing and structured-light delivery}

For semiconductor-dot implementations, the internal transitions addressed by twisted light fall in the $\sim 0.2$--$0.7$~THz band for representative parameters~\cite{APSOS2026}, corresponding to free-space wavelengths of $\sim 0.4$--$1.5$~mm. A diffraction-limited far-field spot at these wavelengths is at least three orders of magnitude larger than typical dot pitches ($\sim 100$~nm) and considerably larger when realistic numerical-aperture constraints of THz optics are accounted for. Closing this wavelength--pitch mismatch is a generic challenge for any THz-driven semiconductor qubit; it is the reason that semiconductor charge qubits and the charge quadrupole qubit~\cite{Friesen2017,Koski2020} have been controlled by gate voltages and microwave-frequency cavity coupling rather than by free-space optics.

Surface plasmon polaritons (SPPs) provide the standard route past the diffraction limit by confining electromagnetic energy at metal--dielectric interfaces~\cite{Maier2007}; adiabatic nanofocusing on conical tips~\cite{Stockman2004} concentrates SPP energy into sub-wavelength volumes. 
In ordinary-metal conical tips, only the azimuthally symmetric \(m=0\) SPP mode reaches the tip end; ordinary-metal nanofocusing therefore does not provide a family of nonzero-OAM modes with selectable sign.
This is a serious limitation because the qubit gate logic requires delivery of a structured near field with the required angular-momentum content $s=l+\sigma$.

The relevant breakthrough comes from work on Weyl-semimetal plasmonics. 
Hofmann and Das Sarma~\cite{Hofmann2016} showed that Weyl-node separation modifies the SPP dispersion: in time-reversal-broken Weyl semimetals, the momentum-space node separation acts as an effective magnetic field for surface plasmons, producing nonreciprocal surface modes.
In a related cylindrical-waveguide study, Peluso, De Martino, Egger, and Buccheri~\cite{Buccheri2025} extended this to nonreciprocal Weyl-semimetal waveguides, demonstrating that OAM is a control parameter for SPP propagation direction with a giant splitting in group velocity between opposite-OAM modes.
Crucially, Peluso et al.~\cite{Peluso2025} show theoretically that SPPs on a magnetic Weyl-semimetal conical tip exhibit OAM-selective nanofocusing: the axion term in the effective electrodynamics modifies the dispersion such that, within the relevant frequency window, all modes with one sign of OAM reach the apex while the opposite-sign modes are radiated before the tip. Reversing the Weyl-node-separation, or anomalous-Hall, vector reverses which OAM sign is focused. 
The broader review of light control with Weyl semimetals~\cite{Guo2023} situates this within the rapidly developing fields of topological photonics and chiral metamaterials.

\subsection{Contribution of this work}

We propose that the combination of (i) generalized-Kohn-theorem  symmetry protection of relative-motion qubit states in parabolic confinement, (ii) twisted-light OAM-driven control of those states, and (iii) Weyl-semimetal-enabled OAM-selective nanofocusing constitutes a coherent architecture for practical quantum information processing. Each of the three pillars rests on an independent body of well-established physics; the architecture coheres because the structured-light character required by the protection mechanism is precisely the structured-light character that the metamaterial channel preserves and that the gate logic exploits. Section~II develops the symmetry-protected encoding via the Kohn theorem. Section~III formulates the twisted-light control channel. Section~IV presents the metamaterial nanofocusing integration layer. Section~V discusses comparison with competing platforms, the experimental-maturity question relative to the charge quadrupole qubit, platform portability across parabolic-confinement systems (cold atoms, trapped ions, levitated nanoparticles), and honest limitations. Section~VI concludes.

\section{Symmetry-protected encoding via the Kohn theorem}

The generalized Kohn theorem, established in its modern form for parabolic 2D confinement \cite{BJH1989,BJ1990,QJ1,QJ2} and absorbed into the standard quantum-dot literature thereafter~\cite{Reimann2002}, states the following. For a system of interacting electrons in a parabolic confining potential, any spatially uniform dipolar perturbation couples only to the center-of-mass coordinate. The internal (relative-motion) spectrum is invariant under such perturbations, independent of the interaction strength or form.

The implication for qubit design is direct. Consider two interacting electrons in a 2D parabolic dot of frequency $\omega_0$ under perpendicular magnetic field $B$. Defining relative coordinate $z = z_1 - z_2$ and center-of-mass $Z = (z_1 + z_2)/2$, the Hamiltonian separates exactly into CM and relative parts~\cite{Rodriguez2025,APSOS2026}. Define computational basis states from two consecutive odd-numbered $|m|$ relative-motion correlation sectors, e.g.
\begin{equation}
|0\rangle \equiv | |m| =3\rangle, \qquad |1\rangle \equiv | |m| =5\rangle
\label{eq:basis}
\end{equation}
though we stress that other odd-numbered pairs could be chosen \cite{APSOS2026} and hence this scheme is remarkably flexible. 
Three consequences follow:

\begin{enumerate}
\item Each definite-$m$ state in a circular dot is rotationally symmetric and satisfies $\langle \mathbf{r}\rangle = 0$. It has no permanent in-plane dipole moment.
\item Any spatially uniform electric-dipole field approximately uniform on the dot scale couples to the center of mass but not to the qubit subspace. The qubit splitting $\hbar\omega_{\text{qubit}}$ is unaffected by this electric-dipole channel in the ideal harmonic model.
\item The leading remaining sensitivity is quadrupolar: fluctuations of the confining curvature shift the levels through their dependence on $\langle r^2 \rangle$, which differs between $|0\rangle$ and $|1\rangle$. The Hellmann--Feynman estimate gives
\begin{equation}
\sigma_{\omega_{\text{qubit}}} = \frac{\mu \omega_0}{\hbar}\,\Delta\langle r^2 \rangle\, \sigma_{\omega_0},
\label{eq:dephasing}
\end{equation}
where $\sigma_{\omega_0}$ is the rms fluctuation of the bare harmonic confinement, $\mu = m^*/2$ is the relative-motion reduced mass, and $\Delta\langle r^2 \rangle = \langle r^2 \rangle_5 - \langle r^2 \rangle_3$ is the difference of expectation values between the two qubit states.
\end{enumerate}

The protection mechanism is distinct from the established alternatives. Transmons achieve charge-noise insensitivity through parameter-regime suppression: charge dispersion is exponentially small at large $E_J/E_C$, but reappears at higher noise amplitudes and constrains the achievable anharmonicity. Spin qubits achieve E-field insensitivity through the absence of a coupling matrix element, with spin--orbit mixing and exchange operations reintroducing sensitivity. Trapped-ion clock-state qubits achieve protection through atomic structure at specific operating points. The charge quadrupole qubit~\cite{Friesen2017,Koski2020} achieves protection through geometric symmetry of three dots. The protection here, by contrast, is a symmetry property of the relative-motion Hilbert space of a single parabolic trap --- a property of the Hilbert-space decomposition rather than a tuning of parameters or geometric arrangement.

\section{Twisted-light control: selection rules and gate operations}

A symmetry-protected encoding raises an immediate question: if uniform dipolar fields cannot reach the qubit, what {\em can}? A twisted-light mode carrying OAM index $l$ and circular polarization helicity $\sigma = \pm 1$ transfers total angular momentum $s = l + \sigma$ per absorbed photon. For the relative-motion correlation sector encoding, the transition selection rule is~\cite{Rodriguez2025,APSOS2026}
\begin{equation}
\Delta |m| = \pm s = \pm(l+\sigma).
\label{eq:selection}
\end{equation}
For $l \neq 0$, the optical field has spatial structure on the dot scale, and its multipole components couple directly to the relative coordinate. For the \(|m|=3\leftrightarrow |m|=5\) qubit defined in Eq.~(1),
the required angular-momentum transfer is \(s=2\). A natural
optical choice is therefore \(l=1\) and \(\sigma=+1\), although
any structured mode satisfying \(l+\sigma=2\) would obey the
same angular-momentum selection rule.

The decisive feature for decoherence is that the same selection rule that enables intentional control suppresses coupling to spatially uniform electric-dipole noise. Environmental electrostatic noise typically has its dominant long-wavelength weight in the $l=0$ component, which couples to the center of mass rather than to the internal qubit. Local quadrupolar noise, however, can contain $s=\pm2$ components and is not removed by the Kohn theorem. The encoding is therefore addressable by twisted photons and simultaneously decoupled from the dominant uniform electric-dipole channel, while leaving a concrete quadrupolar-noise engineering target.
In the analytically solvable Calogero limit of the interaction, 
the relative-motion matrix element for a transition driven by a photon carrying total angular momentum \(s=l+\sigma\)
between $|m| \to |m| + s$ takes the closed form~\cite{APSOS2026}
\begin{equation}
M_{m\to m+s}^{(\alpha)} = \frac{\Gamma\left(\frac{\alpha_m + \alpha_{m+s} + s + 2}{2}\right)}{\sqrt{\Gamma(\alpha_m+1)\Gamma(\alpha_{m+s}+1)}},
\label{eq:matrix}
\end{equation}
with $\alpha_m = \sqrt{m^2 + \alpha}$ and $\alpha$ the dimensionless interaction strength. 
Let \(g_s\) denote the energy prefactor of the resonant twisted-light
coupling after the optical and envelope factors have been included. Then,
within the rotating-wave approximation,
\begin{equation}
\Omega_R = \frac{2|g_s|}{\hbar}\,
\left|M_{3\to5}^{(\alpha)}\right| ,
\end{equation}
up to pulse-shape conventions. Free evolution between pulses generates $R_z$ rotations through the qubit splitting $\hbar\omega_{\text{qubit}} = \hbar\omega(\alpha_5 - \alpha_3) - \hbar\omega_c$, with $\omega = \sqrt{\omega_0^2 + \omega_c^2/4}$ and $\omega_c = eB/m^*$. Together, $\{R_x(\theta), R_z(\phi)\}$ form a universal single-qubit gate set~\cite{APSOS2026}.

The same structured-light channel supports three operations through one physical mechanism: \textsc{Write} (resonant TL pulse implements $R_x$ between sectors), \textsc{Read} (TL spectroscopy of correlation-sensitive lines distinguishes the sectors), and \textsc{Scale} (spatial-light-modulator steering plus vorticity selection addresses many sites without dense on-chip wiring). The selection rules that protect the qubit from environmental noise simultaneously prescribe the required angular-momentum class of structured modes that can address it.

\section{Metamaterial nanofocusing: closing the integration gap}

The two preceding pillars --- generalized Kohn protection and twisted-light control --- require that a structured photon actually reach the qubit at the right length scale. For semiconductor implementations, this is the engineering frontier where recent metamaterial-photonics advances provide a concrete answer.

\subsection{The wavelength--pitch mismatch}

For a representative GaAs dot with $\hbar\omega_0 \sim 3$~meV at $B \sim 3$--$5$~T, the internal transitions addressed by twisted light fall in the $\sim 0.2$--$0.7$~THz band, with free-space wavelengths $\lambda \sim 0.4$--$1.5$~mm. A diffraction-limited far-field spot exceeds typical dot pitches ($\sim 100$~nm) by at least three orders of magnitude, and considerably more for realistic THz numerical apertures. No far-field beam shaping can resolve a single dot in a dense array.

This challenge is generic for THz-driven semiconductor qubits and motivates the use of gate-voltage or cavity-microwave control in competing schemes such as the charge quadrupole qubit~\cite{Friesen2017,Koski2020}. For the present architecture, however, the OAM character of the control channel is intrinsic to the protection logic and cannot be replaced by an unstructured control modality without surrendering the symmetry-protected addressability.

\subsection{OAM-selective SPP nanofocusing on Weyl-semimetal tips}

Surface plasmon polaritons confine electromagnetic energy at metal--dielectric interfaces and can be nanofocused by adiabatic conical-tip geometries~\cite{Maier2007,Stockman2004}. In ordinary-metal conical tips, only one SPP mode reaches the tip end; ordinary-metal nanofocusing therefore does not supply the family of nonzero-OAM near-field channels required by Eq.~(\ref{eq:selection}).
But Peluso et al.'s work ~\cite{Peluso2025} mentioned above, shows that this limitation is removed for magnetic Weyl semimetals and is hence worth restating in full: the axion term in the effective electrodynamics modifies the SPP dispersion such that, within the relevant frequency window, all modes with one selected sign of OAM reach the tip apex while the opposite-sign family is radiated before the apex. Reversing the Weyl-node-separation, or anomalous-Hall, vector reverses the focused OAM sign, providing a route to the $+s$ and $-s$ gate channels.
The result builds on foundational work establishing the modified SPP dispersion of Weyl semimetals~\cite{Hofmann2016}, on cylindrical-waveguide analyses demonstrating OAM as a control parameter for SPP propagation~\cite{Buccheri2025}, and on the conical-tip nanofocusing result of Peluso et al.~\cite{Peluso2025}; it fits within the broader program of topological photonics and chiral metamaterials~\cite{Guo2023}.

\subsection{Implications for the architecture}

For the proposed architecture, this body of work supplies a concrete photonic-integration route. 
A free-space twisted-light beam, coupled into a magnetic Weyl-semimetal conical tip through an appropriate antenna, grating, or near-field coupler, can launch SPPs that nanofocus to a sub-wavelength near-field region overlapping one or a few quantum dots.
The key architectural requirement is that the OAM-selective SPP channel produce, at the dot position, a dominant local cylindrical harmonic with the required total angular momentum $s=l+\sigma$. 
The sign of the focused OAM family is controlled by the Weyl-node-separation, or anomalous-Hall, vector, providing a reconfigurable channel between gate orientations.
Array addressing then becomes plausible: multiple tips with independently set magnetizations can address multiple dots in parallel from a shared incident beam, replacing dense on-chip control wiring with a photonic-metamaterial control bus that realizes the SCALE concept of Section~III.

The Peluso et al.\ proposal has not been experimentally realized, and the materials development of magnetic Weyl semimetals at the relevant length and frequency scales remains an active engineering frontier. However, the route is concrete, the underlying physics is well-grounded in the metamaterial literature on chiral and topological photonics~\cite{Guo2023}, and the OAM-selectivity that distinguishes Weyl-SPP nanofocusing from ordinary-metal SPP nanofocusing is precisely the property that the present architecture requires.

\section{Discussion}

\subsection{Comparison with competing schemes}

Table~\ref{tab:platforms} summarizes the principal charge-noise-protection mechanisms in mature and proposed quantum-computing platforms. The closest direct competitor on the protection goal is the charge quadrupole qubit of Friesen et al.~\cite{Friesen2017}, which shares the goal of immunity to uniform electric-field fluctuations but differs in mechanism (geometric DFS in three dots vs.\ generalized-Kohn-theorem symmetry in one), geometry (triple vs.\ single dot), control channel (gate-voltage / microwave-cavity vs.\ twisted light), particle number (one electron vs.\ two or more), and topological-protection roadmap (none vs.\ $N\geq 3$ Laughlin-quasihole extension).

\begin{table*}[!t]
\centering
\caption{Charge-noise protection mechanisms across major QC platforms.}
\label{tab:platforms}
\renewcommand{\arraystretch}{1.25}
\small
\begin{tabular}{p{3.4cm} p{4.0cm} p{3.4cm} p{5.0cm}}
\toprule
\textbf{Platform} & \textbf{Protection mechanism} & \textbf{Category} & \textbf{Residual channels / limitations} \\
\midrule
Semiconductor charge qubit~\cite{Hayashi2003,Petersson2010} & Sweet-spot operation; $T_2^\ast$ extended from $\sim 1$ to $\sim 7$~ns & Operating-point tuning & 1/$f$ charge noise unmitigated outside the sweet spot \\
Charge quadrupole qubit~\cite{Friesen2017,Koski2020} & DFS via three-dot symmetry; uniform field shifts logical states equally & Geometric DFS & Quadrupolar charge noise; fabrication complexity \\
Singlet--triplet qubit~\cite{Petta2005,Reed2016,Martins2016} & Symmetric exchange at sweet spot; vanishing dipole at SOP & Operating-point tuning & Dephasing away from SOP; nuclear hyperfine \\
Superconducting transmon & Exponentially small charge dispersion at large $E_J/E_C$ & Parameter regime & Flux noise, dielectric loss, weak anharmonicity \\
Semiconductor spin qubit & Spin--E-field coupling absent at first order & Vanishing matrix element & Spin--orbit, exchange-mediated reintroduction \\
Trapped-ion clock state & Hyperfine transition insensitive to magnetic field at operating point & Atomic structure & Motional heating, off-resonant scattering \\
Majorana / topological qubit & Non-local fermion-parity encoding & Topological invariant & Not yet realized in robust scalable form \\
\textbf{TL-controlled correlation-sector qubit (this work)} & \textbf{Generalized-Kohn-theorem decoupling of relative motion from uniform dipolar fields} & \textbf{Symmetry (Hilbert-space)} & \textbf{Quadrupolar curvature noise} \\
\bottomrule
\end{tabular}
\end{table*}

The protection mechanism is therefore categorically distinct from those used by transmons (parameter regime), spin qubits (matrix-element absence), ion clock states (atomic structure), and the charge quadrupole qubit (geometric DFS). It is generalized-Kohn-theorem protection of a many-body internal encoding in a single parabolic trap against spatially uniform electric-dipole noise, controlled by structured photons and integrated through metamaterial nanofocusing.

\subsection{The experimental-maturity question}

The charge quadrupole qubit has been experimentally realized~\cite{Koski2020} while the present scheme remains at the theoretical-proposal stage. The Koski et al.\ experiment achieved three concrete milestones: strong coupling between a superconducting microwave resonator and the electric quadrupole moment of a single electron in a linear triple GaAs/AlGaAs quantum dot, operation at a working point where dipole coupling vanishes by geometric symmetry, and spectroscopic confirmation that coherence is limited by short-range (quadrupolar) charge noise rather than the dipolar channel. These are significant results that establish protection from uniform electric-field fluctuations is achievable in a real semiconductor device.

At the same time, the 2020 experiment did not demonstrate the predicted 10--1000$\times$ improvement in gate fidelity over conventional charge qubits, did not demonstrate two-qubit operations between coupled quadrupole qubits, and characterized the protection mechanism via spectroscopy rather than via the full pulse trajectory of a logical gate. The experiment is best understood as a proof of principle for the protection mechanism, not a completed qubit module.

Both schemes share the same residual decoherence channel. The dominant remaining decoherence in Koski et al.\ is short-range quadrupolar charge noise; the leading dephasing channel in the present scheme, captured by Eq.~(\ref{eq:dephasing}), is fluctuation of the confining curvature. Both arrive at the same engineering frontier. The protection-mechanism question is largely settled in both cases; what remains is whether material improvements (substrate choice, gate symmetry, dynamical decoupling) can suppress the quadrupolar noise floor to fault-tolerant levels. An experimental advance against quadrupolar noise in one scheme is likely to benefit the other.

\subsection{Platform portability}

The charge quadrupole qubit is tied to a specific geometric implementation in gate-defined semiconductor heterostructures. There is no straightforward cold-atom version, ion-trap version, or levitated-particle version of the three-dot geometric DFS. The generalized-Kohn-theorem protection of the present scheme, by contrast, is a generic consequence of parabolic confinement and extends to any platform in which OAM-carrying photons can be focused onto the trap volume.

{\em Cold atoms.} Léonard et al.~\cite{Leonard2023} have realized a small-$N$ Laughlin-type state demonstrating the kind of inter-particle correlations the present scheme exploits. Visible-wavelength OAM beams directly match typical tweezer spacings ($\sim 1$--$5\,\mu$m), so the SCALE mechanism does not require the metamaterial-nanofocusing step needed in the semiconductor implementation. Spatial light modulators with high OAM mode purity are standard cold-atom infrastructure. Combined with proposals for minimal cold-atom quasihole setups~\cite{Umucalilar2023} and FQH preparation paths~\cite{Popp2004,Grusdt2014,Barkeshli2015,Repellin2017,Lacki2025}, the cold-atom platform may reach a complete qubit demonstration of the present architecture on a timescale comparable to or shorter than further development of the charge quadrupole qubit beyond its current spectroscopic stage.

{\em Trapped ions.} Schmiegelow et al.~\cite{IonOAM2016} have demonstrated direct transfer of optical orbital angular momentum to a bound electron in a single trapped ion. Afanasev et al. \cite{Andrei} have  made significant advances in this area. Radial confinement in a Paul trap is parabolic to good approximation, so the Kohn theorem applies. The relevant relative-motion states of a small ion crystal would offer the same symmetry protection, combined with the demonstrated single-shot readout fidelities of mature ion-trap laboratories. 

{\em Levitated nanoparticles and other systems.} Quantum-optomechanics platforms based on optically levitated nanoparticles operate near the harmonic limit, with generalized-Kohn-theorem decoupling between center-of-mass and internal modes assisting in current cooling protocols. FQH droplets in parabolic edge confinement, electron bubbles on liquid helium, and engineered NV-center strain manifolds provide further candidate platforms. The common requirement is parabolic confinement and a noise spectrum dominated by spatial scales larger than the qubit footprint.

The architecture therefore offers an implementation portfolio: four parallel experimental platforms sharing one design language, with the cold-atom path as the likely near-term first demonstration and the metamaterial-enabled semiconductor path as a parallel longer-term route. From a deployment standpoint, this reduces the risk that an unforeseen engineering bottleneck in any single platform invalidates the architecture.

\subsection{Limitations}

Our proposal currently has several limitations, though we expect progress can still be achieved in a short period of time:

First, the Kohn theorem protects against linear (dipolar) fields but not against fluctuations of the confining curvature. Equation~(\ref{eq:dephasing}) shows that the residual quadrupolar dephasing channel, while generically weaker than the dipolar channel it replaces, is not zero. This is the same channel that limits the charge quadrupole qubit~\cite{Kornich2018,Koski2020}; the two architectures converge on the same engineering frontier.

Second, the topological-protection roadmap requires $N \geq 3$ electrons. The $N = 2$ system analyzed here delivers a symmetry-protected single-qubit primitive; true topological order, quasihole braiding, and the associated stronger protection appear only at three or more particles. A 1/$N$-expansion scaffold for the $N = 3$ extension is provided in our companion paper~\cite{APSOS2026} but is not experimentally realized.

Third, the metamaterial nanofocusing pillar relies on theoretically established but experimentally unrealized OAM-selective Weyl-semimetal SPP transport~\cite{Peluso2025}. Materials development of magnetic Weyl semimetals at the relevant length and frequency scales for THz operation remains an active engineering frontier; the cold-atom implementation of the architecture does not require this development, as the wavelength--pitch matching is already favorable in the visible.

We also note that for gate operation, OAM-selective SPP transport is necessary but not
sufficient. The nanofocused near field at the dot must retain a dominant
local cylindrical-harmonic component with the required total angular
momentum \(s=l+\sigma\), and unwanted \(s'\neq s\) components must be
small compared with the qubit transition selectivity. This requirement
should be checked by a near-field electromagnetic simulation and a local
multipole decomposition about the dot center.
\section{Conclusion}
We have proposed an architecture for practical quantum information processing built on three independent physical principles. {\em Protection} comes from the generalized Kohn theorem~\cite{BJH1989}, which decouples relative-motion correlation sectors of a parabolic trap from spatially uniform electric-dipole noise. {\em Control} comes from twisted-light selection rules, which prescribe the angular-momentum class $s=l+\sigma$ needed to address the protected qubit. {\em Integration} comes from OAM-selective metamaterial nanofocusing on magnetic Weyl-semimetal tips~\cite{Peluso2025}, which offers a route to closing the THz wavelength--pitch mismatch in the semiconductor implementation while delivering the structured near-field content required by the gate logic.

The protection mechanism is categorically distinct from the charge-noise suppressions used by competing platforms. The experimental-maturity gap relative to the closest direct competitor, the charge quadrupole qubit~\cite{Friesen2017,Koski2020}, is narrower than the publication record suggests --- both schemes successfully suppress the linear stray-field channel and converge on the same residual quadrupolar engineering frontier. The protection principle is platform-portable across cold atoms, trapped ions, levitated nanoparticles, and semiconductor dots, with cold atoms as the likely near-term first-demonstration path and metamaterial-enabled semiconductors as a parallel longer-term route.

The architecture is a clear motivating application for the metamaterial-photonics agenda. OAM-selective nanofocusing, chiral plasmonics, topological photonics, and Weyl-semimetal optics --- each of these communities has developed capabilities that the present quantum-information architecture requires. The convergence of these capabilities, on a timescale set by the slowest of the three pillars rather than by any single bottleneck, defines a practical path from foundational symmetry theorems to deployable quantum hardware.

\section*{Acknowledgments}

F.J.R.\ and L.Q.\ acknowledge financial support from Project No.\ 5228 Banco de la Rep\'ublica (2025, Colombia) and from Facultad de Ciencias Project No.\ INV-2025-213-3449, Universidad de Los Andes (Colombia).

\bibliographystyle{IEEEtran}

\end{document}